\documentclass[aps,prb,twocolumn,longbibliography,groupedaddress,showpacs,showkeys,citeautoscript,amsfonts,amsmath,amssymb,flushbottom,floatfix]{revtex4-1}
\usepackage{graphicx}
\usepackage{nicefrac}
\usepackage[latin1]{inputenc}
\usepackage[T1]{fontenc}
\usepackage{wasysym}   
\usepackage{float}

\usepackage{color}

\newif\ifshowcolor

\ifshowcolor
\newcommand{\bl}[1]{\textcolor{blue}{#1}}

\else
\newcommand{\bl}[1]{#1}

\fi

\begin{document}
\setlength{\floatsep}{0pt}
\setlength{\intextsep}{0pt}

\newcommand{\bskipdm}{\mskip -3.8\thickmuskip}
\newcommand{\bskiptm}{\mskip -3.0\thickmuskip}
\newcommand{\bskipsm}{\mskip -3.1\thickmuskip}
\newcommand{\bskipssm}{\mskip -3.1\thickmuskip}
\newcommand{\fskipdm}{\mskip 3.8\thickmuskip}
\newcommand{\fskiptm}{\mskip 3.0\thickmuskip}
\newcommand{\fskipsm}{\mskip 3.1\thickmuskip}
\newcommand{\fskipssm}{\mskip 3.1\thickmuskip}
\newcommand{\pint}{\mathop{\mathchoice{-\bskipdm\int}{-\bskiptm\int}{-\bskipsm\int}{-\bskipssm\int}}}
\newcommand{\ds}{\displaystyle}
\newcommand{\D}{\mathrm{d}}
\newcommand{\I}{\mathrm{i}}
\newcommand{\EXP}[1]{\mathrm{e}^{#1}}    

\newcommand{\tpp}{\hat{t}}
\newcommand{\rpp}{\hat{r}}
\newcommand{\zpp}{\hat{z}}
\newcommand{\ppp}{\hat{\varphi}}
\newcommand{\tbb}{\bar{t}}
\newcommand{\rbb}{\bar{r}}
\newcommand{\zbb}{\bar{z}}
\newcommand{\pbb}{\bar{\varphi}}
\newcommand{\lbb}{\bar{\ell}}
\newcommand{\ttt}{\tilde{t}}
\newcommand{\rtt}{\tilde{r}}
\newcommand{\ztt}{\tilde{z}}
\newcommand{\ptt}{\tilde{\varphi}}
\newcommand{\abs}[1]{\left\lvert #1 \right\rvert}
\newcommand{\tvv}{\check{t}}
\newcommand{\xvv}{\check{x}}
\newcommand{\omvv}{\check{\omega}}
\newcommand{\kvv}{\check{k}}
\newcommand{\vvv}{\check{v}}
\newcommand{\lvv}{\check{\ell}}

\newcommand{\abl}[3][\empty]{\frac{\D^{#1} #2}{\D {#3}^{#1}}}
\newcommand{\pabl}[3][\empty]{\frac{\partial^{#1} #2}{\partial {#3}^{#1}}}

\title{Radially falling test particle approaching an evaporating black hole}

\author{Klaus Kassner}
\affiliation{Institut für Physik,
  Otto-von-Guericke-Universität
  39110 Magdeburg, Germany
}

\begin{abstract}
  A simple model for an evaporating non-rotating black hole is
  considered, employing a global time that does not become singular at
  the putative horizon. The dynamics of a test particle falling
  radially towards the center of the black hole is then
  investigated. Contrary to a previous approach, we find that the
  particle may pass the Schwarzschild radius before the black hole has
  gone. \bl{Backreaction effects of Hawking radiation on the
    space-time metric are not considered, rather a purely kinematical
    point of view is taken here.} The importance of choosing
  an appropriate time coordinate when describing physical processes in
  the vicinity of the Schwarzschild radius is \bl{emphasized}. For a
  shrinking black hole, the true event horizon is found to be inside
  the sphere delimited by that radius.
\end{abstract}

\date{May 9, 2018}


\maketitle

\section{Introduction}

In a recent article in this journal,\cite{aste05} Aste and Trautmann
discussed the radial fall of a test particle onto a black hole
evaporating by Hawking radiation. They did so in terms of what they
admit to be a toy model, an approximation the validity of which is, in
their own words, ``dubious at best'' near the event horizon of the
black hole. Nevertheless, they believe their model to be a good
approximation far from the horizon and consider their major
qualitative result to be correct, i.e., that the black hole will
evaporate under an infalling particle or observer before the latter can cross the
horizon. Such a result would call into question that a black hole can
form at all \cite{vachaspati07}, because any piece of infalling matter
should hover above the \bl{nascent} horizon, accumulating time dilation,
while the incipient black hole is already busy evaporating.

The purpose of this paper is to show that a slightly more realistic
approach, still a toy model, but at least one using a time coordinate
that does not become singular at the Schwarzschild radius, leads to a
diametrically opposed prediction: an observer falling towards an
evaporating black hole will normally cross the ``horizon'' without
noticing anything dramatic happening and, unfortunately, hit the
central singularity the same way as with an eternal black hole. Rather
precise timing of the fall, approaching the black hole near the end of
its lifetime, when its mass has gotten small, would be necessary to
avoid hitting it before it evaporates. Moreover, the Schwarzschild
radius of an evaporating black hole loses its property of being an
event horizon. Rather, its behavior bears some similarities to that of
the Hubble sphere in cosmology,\cite{davis04} the locus of galaxies
receding from an observer at the speed of light:\footnote{A light
  signal sent from outside the Hubble sphere may reach us, due the the
  continuing expansion of the latter so it may eventually catch up
  with an inward-running light front. Once that light is inside the
  Hubble sphere, it will reach its center, in
  principle. 
} a light signal sent outward from the Schwarzschild radius or even
from slightly inside it can reach outside observers due to the
shrinking of the radius.

These conclusions seem to be at odds not only with
Ref.~[\onlinecite{aste05}] but also with [\onlinecite{vachaspati07}],
who use a genuine quantum mechanical formalism. However, the
calculations in Ref.~[\onlinecite{vachaspati07}] are based on the
Schwarzschild time coordinate and therefore subject to the
restrictions imposed by its use, even if this does not show up in a
divergence of the calculation. We will discuss the nature and
operational meaning of different time coordinates in Sec.~2.  As we
shall see, time coordinates constructed in an analogous way to the
Schwarzschild time must become infinite for events on a horizon or
one-way membrane for light propagation. If a quantum mechanical
Hamiltonian is chosen conjugate to this time coordinate, it will
obviously not be able to capture any dynamics at the horizon, whether
or not this is signaled by an explicit divergence. Using this time
automatically imposes limitations on the range of the description in a
way not required by some other time coordinates. These restrictions
then may lead to a misinterpretation of mathematically correct results.

\bl{The calculations in this paper are purely classical, i.e., no
  attempt is made at a quantum mechanical model of the
  radiation. There are controversial statements in the literature
  about the effects of backreaction of Hawking radiation. While there
  seems to be general agreement that the emission of Hawking radiation
  begins already during the collapse of a star, i.e., before the
  formation of an event horizon, some authors
  \cite{gerlach76,mersini14,peiming16} believe that the ensuing
  reduction of mass-energy may prevent the formation of a horizon
  altogether, whereas others \cite{parentani94,paranjape09,modak14}
  insist that it is too small to have such a drastic effect. We will
  not tackle these questions which refer to the \emph{dynamics} of
  black-hole formation. Rather, the focus will be on getting the
  \emph{kinematics} right, which, as I argue here, necessitates to avoid
  using a time coordinate that is singular at the apparent horizon. In the
  conclusions, I will return to the question whether we may learn
  something regarding the aforementioned controversy from our results
  as well.  In any case, the Aste-Trautmann calculation does not take
  backreaction effects into account and nevertheless predicts
  evaporation of a black hole from under the feet of an infalling
  observer, so to speak. An attempt to improve a little on this
  calculation therefore need not invoke backreactions either.  }

The remainder of this paper is organized as follows. In
Sec.~\ref{sec:time_in_genrel}, Painlevé-Gullstrand (PG) coordinates
for the metric of an eternal spherically symmetric black hole are
introduced besides the standard Schwarzschild ones. Ways to
operationally realize the synchronies established by these coordinates
are presented and the reason for the singularity of standard
Schwarzschild coordinates at the event horizon is explained. It is
shown that in PG coordinates not only \bl{does} the proper time of a particle 
falling into the black hole remain finite but also the global time.
Section \ref{sec:toy_model} briefly recalls the model of an
evaporating black hole introduced by Aste and Trautmann and gives an
alternative model based \bl{on} a similar construction but now using the PG
time coordinate to describe the decay law due to Hawking
radiation. The equations of motion for a particle falling radially
towards the black hole are given for this model and
simplified. Finally, in Sec.~\ref{sec:fate_infall}, these equations as
well as those describing outgoing light rays are solved
numerically. The black hole does not normally evaporate, before an
infalling object passes the Schwarzschild radius. A summary and
conclusions are given in Sec.~\ref{sec:conclus}.

\section{Time in general relativity}
\label{sec:time_in_genrel}

Our concept of time has been radically altered by the relativity
theories. Just how radical the nature of this change turns out to be,
may not have been as well appreciated as the wide
acceptance of these theories would suggest.

Newton believed in absolute time as an independent aspect of
reality. Newtonian simultaneity is objective. It can be ascertained,
in principle, by the fact that there is no limit to the speed of
signals in Newtonian physics. The motion of the center of mass of an
object will immediately affect its gravitational interaction with a
distant detector.

Einstein first deprived time of its absolute character in special
relativity (SR), where different observers may have different notions
of simultaneity, and later reduced it to a mere coordinate of
four-dimensional spacetime, in general relativity (GR). The only
concept of a \emph{physical} time surviving in GR is that of proper time
which is only local.

Already in his paper introducing SR,\cite{einstein05} Einstein
emphasized that his approach to synchronization of clocks and, hence,
simultaneity, was a \emph{definition}, a fact that he later
accentuated by stating that the constancy of the speed of
light underlying his notion of simultaneity was ``neither a
supposition nor a hypothesis about the nature of light but a
stipulation.''\cite{einstein52}

That is, one \emph{may} define simultaneity via the Einstein
synchronization procedure (using light signals and defining the time
of a distant event as the average between the times of emission and
back reception of a light signal sent from the world line of an
observer to the event and immediately reflected back\footnote{This
  procedure makes the one-way speed of light equal to its round-trip
  velocity.}), but this is in \emph{no
  way compulsory}! There is a (gauge) degree of liberty in
establishing synchrony of distant clocks,\cite{rizzi04a} due to the fact that there
is a finite maximum speed of signal transport. This prevents us from
producing a unique standard of simultaneity of distant events. All we
can distinguish objectively is whether pairs of such events are timelike,
null or spacelike.

Nevertheless, the idea still appears to be ingrained in many minds
that, once we have fixed an observer, what is simultaneous at a
distance for this particular observer is objective, which would mean
that a global ``physical'' time could be established at least for
inertial observers. That this is incorrect often does not seem to be
appreciated, and for a reason. If we extend the proper times of
inertial observers via Einstein synchronization to global time
coordinates for each of them (which is possible in a flat spacetime),
this brings out the equivalence of all inertial systems stated in the
relativity principle and renders the Lorentz symmetry of nature
manifest. Therefore, as long as we are dealing, in SR, with (global) inertial
systems only, Einstein synchronization is preferable over all other
synchronization methods on practical
grounds.

On the other hand, when we are concerned with non-inertial systems,
then different synchronization methods may sometimes be preferable
even in SR. Different standard clocks fixed on the rim of a rotating circular disk run
at the same rate as viewed from an observer at the center of the disk.
Nevertheless, if one synchronizes them around the rim with the
Einstein procedure, a time gap will appear between the clock at the
starting point, at angular position $\varphi$ and the one at angle
$\varphi+2\pi$,\cite{kassner12b} i.e., at the same position after
having performed the synchronization procedure for a full turn.
Setting clocks at fixed radial distance by a light signal from the
center of the disk instead\cite{cranor99} (``central synchronization'')
leads to a simple description of phenomena such as the Sagnac
effect, but gives rise to direction-dependent velocities of
light.\cite{kassner12b}

Obviously, practitioners of general relativity are less prone to falling
into the trap of absolutizing observer-conditioned simultaneity,
knowing that simultaneity at a distance is just a matter of choice of
an appropriate spacelike foliation of spacetime (assuming one exists).
There are clearly different choices, even for a specific
observer. And, of course, this is true in SR as well, which can easily
be given a covariant formulation. The mathematics is that of GR with
vanishing Riemannian curvature. Therefore, the same freedom of
establishing ``sheets of simultaneity'' exists in SR as in
GR. 

\emph{Global} times in GR are just time \emph{coordinates}. Generally,
it is not possible to extend the proper time of an observer to a
global coordinate in a way preserving its property of being the proper
time for local observers at different positions, because that would
lead to coordinate singularities in curved spacetime.\footnote{To
  understand why, imagine that the arclength of a planar curve and a
  local normal coordinate are to be extended from a strip about the
  curve to the whole plane. If the curve is not straight, it will have
  osculating circles of finite radius.  Clearly, is not possible to
  assign a unique set of coordinates to the center of such a circle. The normal
  coordinate at two different arclengths will point to that center, so
  it is not described by a single arclength coordinate. The situation gets worse
  for points even farther away from the curve considered.}  To be able
to interpret the coordinate in question as a time, its surfaces of
constant value should of course be spacelike.

After these preliminaries, consider a standard representation of the
metric of a Schwarzschild black hole,
\begin{align} 
  \D s^2 = \left(1-\frac{r_s}{r}\right) c^2 \D t_{S}^2 -
  \left(1-\frac{r_s}{r}\right)^{-1} \D r^2 -r^2\D\Omega^2\>,
\label{eq:schwarzschild_metric}
\end{align}
where $r_s=2 G M_0/c^2$ is the Schwarzschild radius. $M_0$ is
the mass of the black hole, while $G$ is Newton's
gravitational constant and $c$ is the speed of light. $\D\Omega^2= \D
\vartheta^2 + \sin^2\vartheta \D \varphi^2$ is the squared line
element on the surface of a unit sphere, and we have given the time
coordinate a subscript $S$ to distinguish it from alternative ones.

How is the time coordinate in the metric
\eqref{eq:schwarzschild_metric} constructed? It is not the proper time
of any observer stationary within the metric, except for those at
infinity. This can be seen immediately via calculation of the proper
time interval $d\tau$ for such an observer
($\D s^2 = c^2 \,\D\tau^2$). Assuming $\D r=\D\vartheta=\D\varphi=0$,
we find
\begin{equation}
\D\tau = \sqrt{1-\frac{r_s}{r}} \D t_S\>.
\end{equation}
This is not a total differential, so this proper time cannot be integrated
to provide us with a global time coordinate. However, local clocks at a
radial distance $r$ from the center can be made to run fast by a factor of
$\left(1-\frac{r_s}{r}\right)^{-1/2}$ with respect to local standard
clocks.\footnote{Standard clocks are running at the rate of their
  proper time. The clocks of the Global Positioning System are
  nonstandard clocks similar to the ones showing Schwarzschild time,
  except that they run slow with respect to a local standard clock, in
  order to keep them synchronous with clocks on Earth, not at
  infinity.} Presently, we will not consider the case $r<r_s$.
For a discussion of how observers may determine that they are at rest
within the spacetime described by the metric
\eqref{eq:schwarzschild_metric}, see Ref.~[\onlinecite{misner73}].
Once nonstandard clocks running at the same rate have been
established,\footnote{This rate is identical to that of a (standard)
  master clock at infinity.} we only have to synchronize them, i.e.,
to fix the offset of each clock. The synchronization procedure leading
to the time $t_S$ (up to a constant offset) is Einstein
synchronization, based on the requirement that the time a light signal
takes to travel from an observer $A$ to an observer $B$ is the same as
the time it takes from $B$ to $A$. Note that this is valid with the
Schwarzschild time, even though the coordinate speed of light is
neither equal to $c$ nor constant. Light propagation is described by
$\D s^2=0$, which for $\D\vartheta=\D\varphi=0$ leads to
$\D r/\D t_S=\pm\left(1-r_s/r\right)c$, and the coordinate velocities
of light in the $\vartheta$ and $\varphi$ directions also just change
sign when switching from the forward to the backward direction:
$r\,\D\vartheta/\D t_S=\pm\sqrt{1-r_s/r}\, c$,
$r\sin\vartheta\,\D\varphi/\D t_S=\pm\sqrt{1-r_s/r}\, c$. All that
matters for Einstein synchronization to work, is that the speed of
light is the same in both directions at any point along a spatial
path. This is guaranteed for any diagonal metric with time-independent
coefficients, i.e., a static metric. If the metric is only diagonal,
Einstein synchronization is normally still possible locally, i.e., for
close-by clocks, because the property that the vacuum speed of light does not
depend on the direction along a path is still satisfied on any one
of its infinitesimal pieces. But in the case where we do not have
fixed time-dilation factors that depend on position only, clocks will
have to be constantly resynchronized with their neighbours, in order
to operationally construct the time coordinate. It is not sufficient
to just synchronize them once and let them run at a fixed prescribed
rate.

The spatial coordinate $r=r_s$ describes the \emph{event horizon},
whenever it is part of the solution, i.e., when the whole spherically
symmetric mass distribution, outside of which the metric
\eqref{eq:schwarzschild_metric} holds, does not extend beyond
$r_s$. Let us now ask what time coordinates we may assign to events on
the horizon operationally via Einstein synchronization. Light cannot
return to a distant observer from the horizon. Stated in a highbrow
way, it takes an \emph{infinite} time to return. Since by definition the time
to go to the horizon must be the same as the time to return, both
times must be infinite. So the only time coordinate possible for the
horizon is $\infty$. There are no events at finite time on it, even
for an eternal black hole.
This is just another way of
expressing the fact that Schwarzschild coordinates become singular at
the horizon.

Note that we may generalize this observation: Any time-orthogonal
coordinate system describing a spacetime, in which
coordinate stationary observers stay outside an event horizon that may
be present, will have to assign infinite time to the horizon, so the
coordinates must be singular there.\footnote{Kruskal-Szekeres
  coordinates escape this conclusion by the fact that a coordinate
  stationary observer necessarily crosses the horizon.}

Folklore has it that an observer, starting at a finite distance from a
black hole and falling towards it, will reach the event horizon in
finite proper time but take an infinite amount of time to reach it
``from the perspective of a distant observer''. The first of these two
statements is certainly true and also has a pretty clear physical
meaning, because proper time is something the infaller can read off
his standard clock. The second statement refers to Schwarzschild time
and is also correct, if properly understood. Unfortunately, it is
often misunderstood, then leading to misconceptions such as that the
infalling observer may, just before she crosses the horizon, see the
whole future of the universe or else that the collapse of a star will
halt due to time dilation, before matter can cross the ``critical
radius''.\cite{spivey15}

While the first of these beliefs is simply
wrong,\cite{grib09} the second may be traced to
mistaking the Schwarzschild time for a ``coordinate independent
physical quantity'',\cite{spivey15} an idea that is fully
antithetical to GR, in which the Schwarzschild time merely \emph{is} a
coordinate.  Its diverging at the horizon does \emph{not} mean
that the temporal end of the universe is approached.  The
event in spacetime corresponding to the arrival of an infaller at the
horizon is in no way singular. There is time after and there is space
around.

For an eternal black hole, described by the Schwarz\-schild metric
\eqref{eq:schwarzschild_metric}, it is easy to see that $t_S$ cannot
be a physical time near the event horizon. First, isotropy and
invariance under time translations show that all points on the horizon
are equivalent. Therefore, we expect events to exist on the horizon at
all physical times (this may be taken as the \emph{meaning} of eternal
existence).  But a Kruskal-Szekeres diagram\footnote{See, for example,  
  \protect\url{https://en.wikipedia.org/wiki/Kruskal-Szekeres_coordinates}.}
readily shows that, if we exclude the origin (where the behavior of the Schwarzschild
time is similar to that of the angular polar coordinate near the origin
of a Euclidean coordinate system, in being undefined and arbitrary)
the horizon has only one Schwarzschild time, viz. $t_S=\infty$, even
though it occupies an infinite 3D submanifold of
spacetime.\footnote{There is also an \emph{antihorizon} in the
  diagram, having $t_S=-\infty$. It is the times between these limits
  $-\infty$ and $\infty$ that are missing from the horizons.}
Therefore, interpreting the Schwarzschild time as a physical time
meets some difficulties on the horizon and in its immediate
neighbourhood.


Moreover, the statement that the infaller takes infinite time as seen
from a distant observer is largely devoid of physical meaning. What \bl{it}
means is essentially that it is possible to establish a simultaneity
relation so that the time coordinate of the distant observer that is
simultaneous with the event of the infaller touching the horizon is
infinite. But this again is just a statement about coordinates. There
may be, and as we shall see, there are, choices of time coordinates so
that the event of the infaller on the horizon is simultaneous with a
\emph{finite time} of the distant observer. Translated to the sloppy
language of folklore, this would say that the infaller reaches the
horizon in a finite time from the perspective of a distant observer.
This is equally correct as the original statement, it just refers to a
different time coordinate, that, incidentally, \bl{may become} equal to the
proper time of a stationary observer at sufficiently large $r$ the
same way as the Schwarzschild time does.

What residual of physical meaning we can assign to the fact that the
Schwarzschild time diverges on approach \bl{to} the horizon is that if an
event outside the horizon is in timelike relationship to an observer
just crossing the horizon, it must be in her causal past. Differently
stated, events outside the horizon that are not in the causal past of
an event on the horizon, must be spacelike with respect to it, they
cannot be in its timelike causal future. But this is obvious from the
fact that the timelike future of any event \emph{on} the horizon is
only inside it, ending in the singularity.

A major disadvantage of the Schwarzschild time coordinate is that it
does not give us, without calculation, a time value, beyond which we
know any signal emitted by the distant observer to be unable to reach
an infalling observer. According to Schwarzschild simultaneity, any
external signal is sent before the horizon is reached by the hapless
infaller, so we always have to do a calculation to determine whether
it may or may not reach her.

Let us consider a different form of the Schwarzschild metric:
\begin{align}
 \D s^2 = \left(1-\frac{r_s}{r}\right) c^2 \D t^2 - 2\sqrt{\frac{r_s}{r}} \,c \,\D t \D r - \D r^2 -r^2 \D\Omega^2\>.
\label{eq:schwarzschild_metric_gp}
\end{align}
These are Painlevé-Gullstrand
coordinates\cite{painleve21,gullstrand22} and they are known to be
continuous across the horizon. The spatial coordinates are the same as
in \eqref{eq:schwarzschild_metric}, the time coordinate $t$, to which
I will refer as the Painlevé-Gullstrand or PG time, is related to the
Schwarzschild time by
\begin{align}
c\, t = c\, t_{S}+2\sqrt{r r_s} - r_s \ln\frac{\sqrt{r}+\sqrt{r_s}}{\sqrt{r}-\sqrt{r_s}}+c\, t_R\>,
\label{eq:coordinate_trafo_gptime}
\end{align}
where $t_R$ is an arbitrary constant that can be chosen to make the
two times equal at some fixed radius $R$.\footnote{We will
  assume this from now on, i.e., we set $c\, t_R= r_s
  \ln\left[(\sqrt{R}+\sqrt{r_s})/(\sqrt{R}-\sqrt{r_s})\right]-2\sqrt{R
    r_s}\,$.}  Since the two metrics \eqref{eq:schwarzschild_metric}
and \eqref{eq:schwarzschild_metric_gp} are related by a coordinate
transformation, they describe the same patch of spacetime in their
common domain. The coordinate transformation
\eqref{eq:coordinate_trafo_gptime} is what may be called a
synchronization transformation, because the two times $t$ and $t_S$
run at the same rate for a coordinate stationary observer at $r$,
their only difference being a fixed albeit $r$ dependent offset. Not
every function of $r$ is eligible as an offset, i.e., will lead to an
acceptable new time coordinate. If we write 
\begin{align}
c\,\D t = c\,\D t_{S}+ f'(r) \D r
\end{align}
and assume $\D t_{S}$ to be the positive time interval light takes
to cover the radial distance $\D r$, then $\D t$ must be positive as
well. Since
\begin{align}
c\,\D t_{S} = \left(1-\frac{r_s}{r}\right)^{-1}\abs{\D r}
\end{align}
outside the horizon ($\D s^2=0$, $\D\vartheta=\D\varphi=0$), we
obtain, for $r>r_s$ the inequality
\begin{align}
\abs{f'(r)}<  \left(1-\frac{r_s}{r}\right)^{-1}
\end{align}
as a condition for legitimate synchronization transformations. From
\eqref{eq:coordinate_trafo_gptime}, we find
\begin{align}
f'(r) = \sqrt{\frac{r_s}{r}} \left(1-\frac{r_s}{r}\right)^{-1}\>,
\end{align}
so the inequality is obviously satisfied for $r>r_s$, demonstrating
that $t$ is a legitimate time coordinate outside the horizon,
preserving the time ordering of the Schwarzschild time. \emph{Inside}
the horizon, the continuity of the PG time across
$r=r_s$ suggests that it is a \emph{more} acceptable time coordinate
than the Schwarzschild time $t_S$. In fact, it is known that inside
the horizon $t_S$ takes on the signature of a spatial coordinate
(while the pre\-factor of $\D r$ in \eqref{eq:schwarzschild_metric}
becomes positive, suggesting time like nature), so it is not a 
trustworthy temporal coordinate anymore.

Sometimes the metric \eqref{eq:schwarzschild_metric_gp} is considered
describing a frame of freely falling observers (\emph{rain frame}),
because the local time axis of the coordinate system is parallel to
that of such an observer described in local Minkowski
coordinates. However, this is rather a matter of interpretation than
one of fact. The coordinate stationary observers of the
Painlevé-Gullstrand form of the metric are precisely the same as those
of its Schwarzschild form.

It is useful to have a look at a radially freely falling particle in
this metric. Equations of motion can be obtained from the Lagrangian
$L=\left(\dfrac{\D s}{\D \tau}\right)^2$, and setting
$\D\vartheta=\D\varphi=0$, we need only two of them. Denoting
derivatives with respect to the proper time of our observer by a dot,
we may use the fact that $t$ is a cyclic coordinate and the definition
of $L$ itself:
\begin{align}
\frac12 \frac{\partial L}{\partial\dot{t}} &= \left(1-\frac{r_s}{r}\right) c\dot{t}
 - \sqrt{\frac{r_s}{r}} \,\dot{r} = A = \text{const.}\>,
\label{eq:t_PG_cyclic}
\\
L &= c^2 = \left(1-\frac{r_s}{r}\right) c^2\dot{t}^2 
- 2 \sqrt{\frac{r_s}{r}}  c\dot{t} \dot{r} - \dot{r}^2 \>.
\label{eq:L_PG_stat}
\end{align}
Requiring the particle to have a kinetic energy that would put it in
a coordinate stationary state at infinity, i.e., $\lim_{r\to\infty}
\dot r =0$, we can determine $A$:
\begin{align}
 c^2 = c^2 \dot t^2 \big\vert_{r=\infty} \quad&\Rightarrow \quad \dot t\big\vert_{r=\infty}= 1 
\nonumber\\
A = c\dot t\big\vert_{r=\infty} \quad &\Rightarrow \quad A= c\>.
\end{align}
We may then insert
$\dot{r}=\sqrt{\dfrac{r}{r_s}}\left[\left(1-\dfrac{r_s}{r}\right)c\dot t-c\right]$
from \eqref{eq:t_PG_cyclic} into Eq.~\eqref{eq:L_PG_stat}, which
reduces to a quadratic equation for $\dot t$:
\begin{align}
\dot{t}^2 -\frac{2}{1-r_s/r}\dot{t} + \frac{1+r_s/r}{1-r_s/r} = 0\>
\end{align}
with the solutions
\begin{align}
\dot{t} = \begin{cases} 1 & \text{for } \dot{r}<0
\\
 \dfrac{1+r_s/r}{1-r_s/r} & \text{for } \dot{r}>0
\end{cases}\>.
\label{eq:tdot_falling}
\end{align}
This immediately suggests an operational procedure for clock
synchronization in the Painlevé-Gullstrand frame of reference.  Local
stationary clocks must be nonstandard, as they must run at the same
rate as Schwarzschild clocks ($\partial t/\partial t_S=1$). Then to
synchronize these clocks along a radial line, a standard clock,
initialized to the time of, say an observer at $R$, should be tossed
towards the center from that observer's position with initial velocity
$\dot r = -\sqrt{\dfrac{r_s}{R}} c$ (which turns out to be the same as
$\dfrac{\D r}{\D t}$ here).\footnote{This speed corresponds to the one
  an object falling from rest at infinity would have at $R$.}  Each
stationary clock passed by the falling clock should be set to the time
displayed by the latter the moment of its passage. This way all of
them will be synchronized with the time $t$ of the observer at $R$,
because with these initial conditions $\dot t=1$ according to
Eq.~\eqref{eq:tdot_falling}. Clocks on a shell with fixed radius $r$
may be synchronized with one already set at this radius via Einstein
synchronization, providing the light path is kept at constant $r\>$
(or else by having a clock fall from radius $R$ at each value of
$\vartheta$ and $\varphi$).\footnote{The light could be guided on the
  shell inside glass fibers, because for Einstein synchronization to
  work, we do not need the vacuum speed of light. All that is
  necessary is that the speed of the signal along the forward and
  backward directions of the fiber is the same.}  Note that clocks at
$r>R$ cannot be synchronized with those at $R$ by throwing a clock
\emph{upward} (i.e., in the direction of increasing $r$), because that
does not result in $\dot t=1$. Instead, a clock has to be dropped from
$r>R$, with the observer at $R$ noting its time on passing and sending
the difference between his local time and the time noted to the
observer at larger radius, who can then adjust his clock. This way the
global time $t$ could be operationally realized for all $r>r_s$. For
$r<r_s$, there are no stationary observers anymore, but keeping a
dense stream of falling clocks, we could imagine the time coordinate
to be implemented even inside the horizon, to be read off by anyone
who ever ventures to go there.

Let us now consider, for reference in the next section, the radial
fall towards the horizon of a particle or an observer starting from rest at $r=r_0$.
The equations of motion are still given by
\eqref{eq:t_PG_cyclic} and \eqref{eq:L_PG_stat} but the constant $A$
must be determined anew. We first note that combining
\eqref{eq:t_PG_cyclic} and \eqref{eq:L_PG_stat} we obtain
\begin{align}
A^2 = \dot{r}^2 + \left(1-\frac{r_s}{r}\right) c^2\>,
\label{eq:A2_compact}
\end{align}
so the requirement $\dot{r}\big\vert_{r=r_0}=0$ leads to 
\begin{align}
A^2 = \left(1-\frac{r_s}{r_0}\right) c^2\>.
\label{eq:A2_r0}
\end{align}
Plugging this back into \eqref{eq:A2_compact} we find
\begin{align}
\dot{r} = -\sqrt{\frac{r_s}{r}-\frac{r_s}{r_0}} c\>,
\label{eq:veloc_eq}
\end{align}
whereas $\dot t$ is determined by \bl{(using $A>0$, from Eq.~\eqref{eq:t_PG_cyclic})}
\begin{align}
\left(1-\frac{r_s}{r}\right)\dot{t} = \sqrt{\frac{r_s}{r}} \dot{r} + \sqrt{1-\frac{r_s}{r_0}} c \>.
\label{eq:t_PG_eq}
\end{align}
As was done in Ref.~[\onlinecite{aste05}], we introduce a parameter $\eta$ setting
\begin{align}
r = \frac{r_0}{2}\left(1+\cos\eta\right) = r_0 \cos^2\frac{\eta}{2}\>.
\label{eq:def_r_of_eta}
\end{align}
$\eta=0$ corresponds to the initial position, while $\eta=\pi$ means
that the singularity is being hit. Using \eqref{eq:def_r_of_eta} in
\eqref{eq:veloc_eq}, we obtain an equation for the proper time
$\tau(\eta)$ (the inverse of $\dot\eta$ is the $\eta$ derivative of
$\tau$ and the equation for $\tau'(\eta)$ can be directly integrated):
\begin{align}
c \tau(\eta) = \sqrt{\frac{r_0^3}{4 r_s}}\left(\eta+\sin\eta\right)\>.
\end{align}
The result agrees with that of Ref.~[\onlinecite{aste05}], as it must,
given that $\tau$ was set equal to zero for $\eta=0$ in both
calculations. Moreover, we can also find the global time $t(\eta)$,
from \eqref{eq:t_PG_eq}, using $\dot t =
t'(\eta)/\tau'(\eta)$. Solving this differential equation involves a
slightly demanding integral but can be done analytically exactly, which
provides
\begin{align}
  c t(\eta) &= r_s \sqrt{\frac{r_0}{r_s}-1}
  \left[\eta+\frac{r_0}{2r_s}\left(\eta+\sin\eta\right)\right] \nonumber\\
 & \hspace*{2mm}   + 2\sqrt{r_s r_0}\left(\cos\frac{\eta}{2} -1\right) \nonumber\\
  & \hspace*{2mm} +
  2r_s\ln\frac{\sqrt{1-\frac{r_s}{r_0}}\left(1+\cos\frac{\eta}{2}\right)+\left(1+\sqrt{\frac{r_s}{r_0}}\right)
    \sin\frac{\eta}{2}}{\sqrt{1-\frac{r_s}{r_0}}\left(1+\cos\frac{\eta}{2}\right)+\left(1-\sqrt{\frac{r_s}{r_0}}\right)
    \sin\frac{\eta}{2}}\>.
\end{align}
It is possible though somewhat tedious to verify that the relationship
between this result and the analogous formula for the Schwarzschild
time of Ref.~[\onlinecite{aste05}] is precisely given by
Eq.~\eqref{eq:coordinate_trafo_gptime} with $R=r_0$.  

The horizon is described by $\cos\frac{\eta}{2} =
\sqrt{\frac{r_s}{r_0}}$, so it is reached at
\begin{align}
  c\tau_H = \sqrt{\frac{r_0^3}{r_s}}\left(\arccos\sqrt{\frac{r_s}{r_0}}
  +\sqrt{\frac{r_s}{r_0}}\sqrt{1-\frac{r_s}{r_0}}\right)\>,
  \label{eq:tauH}
\end{align}
which agrees with Ref.~[\onlinecite{aste05}]. In terms of the global
time coordinate, the horizon is attained at
\begin{align}
c t_H &= r_s  \sqrt{\frac{r_0}{r_s}-1}\left(2+\frac{r_0}{r_s}\right)\arccos\sqrt{\frac{r_s}{r_0}}
+ r_0+r_s\nonumber\\  & \hspace*{2mm}-2\sqrt{r_sr_0}
        +2r_s\ln\left(1+\sqrt{\frac{r_s}{r_0}}\right)\>,
        \label{eq:tH}
\end{align}
a perfectly finite time. Moreover, there is no problem determining
both the proper and coordinate times of the observer hitting the
central singularity. We just have to set $\eta=\pi$ and obtain
\begin{align}
  c  \tau_C &= \sqrt{\frac{r_0^3}{4 r_s}} \pi\>,
              \label{eq:tauC}
  \\
c  t_C &=r_s
  \sqrt{\frac{r_0}{r_s}-1}\left(1+\frac{r_0}{2r_s}\right)\pi-2\sqrt{r_sr_0}
  \nonumber\\
  & \hspace*{2mm} +2r_s\ln\frac{1+\sqrt{\frac{r_s}{r_0}}+
    \sqrt{1-\frac{r_s}{r_0}}}{1-\sqrt{\frac{r_s}{r_0}}+\sqrt{1-\frac{r_s}{r_0}}}\>,\hspace*{2cm}
    \label{eq:tC}
\end{align}
and the Painlevé-Gullstrand time remains finite for this event as
well. We can now see one of the advantages of the PG time over the
Schwarzschild one. $t_H$ is simultaneous with the observer hitting the
horizon, $t_C$ simultaneous with her hitting the singularity. Hence,
it is clear that no signal sent by an outside observer \emph{after}
$t_H$ can ever be answered by the infaller and no signal sent after
$t_C$ can ever reach her. The corresponding Schwarzschild times, which
are also finite, must be estimated from solutions of the equation of
motion for a signal chasing the infalling observer or particle. For
the PG times, it is possible to just read them off the description of
the infaller's trajectory.

\section{Toy model of evaporating black hole}
\label{sec:toy_model}

In Ref.~[\onlinecite{aste05}], the authors give, as a simplified model
of an evaporating black hole,
\begin{align} 
  \D s^2 = \left(1-\frac{r_s(t)}{r}\right) c^2 \D t^2 -
  \left(1-\frac{r_s(t)}{r}\right)^{-1} \D r^2 -r^2\D\Omega^2\>,
\label{eq:schwarzschild_evap}
\end{align}
where I will later refer to this time as $t= t_{S}$ and 
\begin{align}
r_s(t) &= \begin{cases}k'\left(t_0-t\right)^{1/3} &\text{for } t\le t_0
\\
0 &\text{for } t> t_0
\end{cases}
\>.
\label{eq:decay_hawk}
\end{align}
This is the time dependence a distant observer would infer from
the relationship for Hawking radiation emitted by a macroscopic black
hole. The temperature of a black hole is, in this limit, inversely
proportional to its mass\cite{opatrny12}
\begin{align}
  T_{\text{BH}}(M) &=\frac{\hbar c^3}{8 \pi G k_B M}\>,
  \label{eq:hawking_temp}
\end{align}
where Planck's constant (divided by $2\pi$) and Boltzmann's constant
appear in standard notations. The thermal radiation of a black body
at this temperature is proportional to $T_{\text{BH}}^4$ but also to
  the surface of the black hole, which goes as $r_S^2 \propto M^2
  \propto T_{\text{BH}}^{-2}$, so the total power output of a black
    hole due to Hawking ratdiation behaves as $T_{\text{BH}}^{2}\propto
      M^{-2}$:
\begin{align}
P_{\text{BH}}(M) &=\frac{\hbar c^6}{240\cdot 64 \pi G^2} M^{-2}
\end{align}
leading to
\begin{align}
\abl{M}{t} &=-\frac{\hbar c^4}{15360 \pi G^2 M^{2}}
\end{align}
and $M^3(t) = \tilde{k} \left(t_0-t\right)$, where 
\begin{align}
  t_0= 5120 M_0^3 \frac{\pi G^2}{\hbar c^4}
  \label{eq:lifetime_BH}
\end{align}
is the lifetime of the black hole and $\tilde{k}=\hbar c^4/(5120 \pi
G^2)$. $k'$ is then just $2G\tilde{k}^{1/3}/c^2$. In
Ref.~[\onlinecite{hiscock81a}], it is argued that the dependency
\eqref{eq:decay_hawk}, based on a fixed-background calculation, cannot
hold down to mass zero, so the functional law must be modified near
$t=t_0$. As we shall see, our calculations suggest a similar
conclusion.

Instead of \eqref{eq:schwarzschild_evap}, I propose the following toy
model for the metric of an evaporating black hole:
\begin{align}
  \D s^2 &= \left(1-\frac{r_s(t)}{r}\right) c^2 \D t^2 - 2\sqrt{\frac{r_s(t)}{r}} \,c \,\D t \D r  \nonumber\\
  &\hspace*{2mm} - \D r^2 -r^2 \D\Omega^2\>,
\label{eq:gp_metric_evap}
\end{align}
with $r_s(t)$ again given by \eqref{eq:decay_hawk}.

Two remarks are in order: First, at some sufficiently large distance
from the black hole, the observed decay law will be the same as in the
Aste-Trautmann model, so this new model is as compatible with Hawking
radiation as the former. Second, the two models are no longer related
to each other by a coordinate transformation, although they ``almost''
are as long as the time dependence of $r_s(t)$ is weak. Nevertheless,
with $r_s$ time dependent, they describe two different physical
situations.

Both models are less realistic than the Vaidya model of
Ref.~[\onlinecite{hiscock81a}] in the following respect: If we
consider the metric of either model at fixed global time and observers
at different radii, the central mass ``seen'' by these observers will
be the same. However, the evaporated energy moves outward so the mass
should increase, if $r$ is increased at \emph{fixed} time,
because a mass shell that has already passed an inner observer
will still be inside the sphere on which a more outward observer is
sitting, who therefore will ``see'' a bigger mass. In the Vaidya
metric, the time coordinate is null, so moving outward at fixed time
means moving outward with the speed of light, staying on the surface
of a volume containing a fixed mass, if we assume the radiation to
move at the speed of light. In this respect, the Vaidya metric is
consistent, while models \eqref{eq:schwarzschild_evap} and
\eqref{eq:gp_metric_evap} are not. 
But they are more easily interpreted, both
having a time-like time coordinate (for $r>r_s$).

This is at least true for the model \eqref{eq:gp_metric_evap} proposed
here, whereas the Aste-Trautmann model has a somewhat severe
conceptual problem. The time coordinate $t_S$ of
\eqref{eq:schwarzschild_evap} has no meaning for $r<r_s(t_S)$, because
it does not establish a simultaneity relationship between the outside
and the inside of the Schwarzschild radius. However, our causal
picture of Hawking radiation is that it transports energy from inside
the volume delimited by $r_s$ to its outside. Now, given the temporal
law $r_s(t_S)$, how is a mass element inside $r_s$ to ``know'' what
time it should ``tunnel'' outside (to keep the law going, so to
speak)?\bl{\protect\cite{modak14} \hspace*{-1mm}} The singular nature of $t_S$ forbids
relating the disappearance of mass inside to its appearance
outside. This means that besides the weak point already mentioned,
consisting in a temporal law of the form $r_s=r_s(t_S)$, the model
even does not give a meaning to the law in the region of space where
the radiation may be thought to originate!  This is different in the
case of the Vaidya metric and for the suggestion made here.  In both
of these cases at least the phenomenological temporal course of the
dynamics is clear, including the relationship between events inside
and outside the ``horizon''.

Next, we wish to describe the dynamics of a test particle falling
radially towards the black hole.  To obtain equations of motion, we use the
effective Lagrangian
\begin{align}
  \mathcal{L}&=\left(1-\frac{r_s(t)}{r}\right) c^2 \dot{t}^2
               - 2\sqrt{\frac{r_s(t)}{r}} \,c \,\dot{t} \dot{r}-\dot{r}^2\>.
\end{align}
The Euler-Lagrange equations
$\ds\abl{}{\tau}\pabl{\mathcal{L}}{\dot{q}}-\pabl{\mathcal{L}}{q}=0$
($q=t,\>r$) then read ($\ds r_{s,t}\equiv \pabl{r_s}{t}$)
\begin{align}
  \left(1-\frac{r_s}{r}\right) c\ddot{t} &=\sqrt{\frac{r_s}{r}}\left(\ddot{r}
                                           -\frac{\dot{r}^2}{2r}\right)
                                           -\frac{r_s}{r^2} \,c\dot{t}\,\dot{r}+
                                           \frac{r_{s,t}}{2r} \,c \dot{t}^2\>,
\label{eq:ddot_t}\\
  \ddot{r} &= -\sqrt{\frac{r_s}{r}}c\ddot{t} - \frac{r_{s,t}}{2r^2} \,c^2 \dot{t}^2 -
             \frac{r_{s,t}}{2\sqrt{r_s r}} \,c \dot{t}^2\>,
\label{eq:ddot_r}
\end{align}
and from $\mathcal{L} = (\D s/\D\tau)^2 = c^2$, we obtain a constant
of motion (the modulus of the four-velocity)
\begin{align}
  c^2 = \left(1-\frac{r_s(t)}{r}\right) c^2 \dot{t}^2
  - 2\sqrt{\frac{r_s(t)}{r}} \,c \,\dot{t} \dot{r}-\dot{r}^2\>.
  \label{eq:fourvel}
\end{align}
Inserting \eqref{eq:ddot_t} into \eqref{eq:ddot_r} to eliminate
$\ddot{t}$ and then replacing $\dot{t}$ (in the terms not containing
$r_{s,t}$) with the help of \eqref{eq:fourvel}, we may considerably
simplify these equations to obtain
\begin{align}
  \ddot{r} = -\frac{r_s(t)}{2 r^2} c^2 - \frac{1}{2\sqrt{r_s r}} r_{s,t}\, c \dot{t}^2\>.
\label{eq:simil_newton}
\end{align}
Obviously, for $r_{s,t}=0$, this has the form of Newton's equation of
motion ($r_s=2 G M_0/c^2$ in that case), with Newton's time replaced
by proper time, an expected result.

Equation \eqref{eq:fourvel} is a quadratic equation for $\dot{t}$ 
\begin{align}
  \dot{t} &=\frac{1}{1-r_s/r}\left(\sqrt{\frac{r_s}{r}}\,\frac{\dot{r}}{c}
            \pm \sqrt{1-\frac{r_s}{r}+\frac{\dot{r}^2}{c²}}\right)\>,
\end{align}
and for an infalling particle ($\dot{r}<0$), the only positive
solution is the one with the plus sign. To remove the prefactor that
diverges at $r=r_s$, we expand the numerator and denominator by the
factor
$\sqrt{\frac{r_s}{r}}\,\frac{\dot{r}}{c}
-\sqrt{1-\frac{r_s}{r}+\frac{\dot{r}^2}{c²}}$
each and find
\begin{align}
  \dot{t} &= \frac{1+\dot{r}^2/c^2}{\sqrt{1-\frac{r_s}{r}+\frac{\dot{r}^2}{c²}}
            + \frac{\abs{\dot{r}}}{c} \sqrt{\frac{r_s}{r}}}\>,
\label{eq:dot_t}
\end{align}
an expression that manifestly remains finite at $r=r_s$. Note that for
$r_s=0$ (i.e., $t>t_0$), \eqref{eq:dot_t} reduces to
\begin{align}
  \dot{t} &= \sqrt{1+\frac{\dot{r}^2}{c^2}} = \sqrt{1+\frac{1}{c^2}\left(\abl{r}{t}\right)^2 \,\dot{t}^2}\>,
\end{align}
implying
\begin{align}
  \abl{\tau}{t} &= \sqrt{1-\frac{\left(\abl{r}{t}\right)^2}{c^2}}\>,
\end{align}
which is the standard result for a particle moving in Minkowski
spacetime (i.e., after the black hole has completely evaporated).

Equations~\eqref{eq:simil_newton} and \eqref{eq:dot_t} are solved
numerically for the set of variables $r$, $v=\dot{r}$, and $t$. In the
particular cases, where integration has to be done beyond $t=t_0$, the
equations with $r_s\ne 0$ are solved up to $t=t_0-\varepsilon$, with a
small value of $\varepsilon$ to avoid the singularity of $r_{s,t}$ at
$t=t_0$ and $r_s$ is set equal to $0$ afterwards. $\varepsilon$ is
varied (made smaller) until the resulting final velocity of the
particle becomes independent of it. This procedure is necessary,
because the used 4th-order Runge-Kutta solver\cite{press86} with step
size adaptation will otherwise halt trying to resolve the (integrable)
singularity of $\ddot{r}$ at $t=t_0$.

In addition, let us consider and solve the equations of motion for an
outgoing light ray in the vicinity of $r=r_s$. Using
\begin{align}
 0 &= \left(\abl{s}{t}\right)^2 =  \left(1-\frac{r_s(t)}{r}\right) c^2 
  - 2\sqrt{\frac{r_s(t)}{r}} \,c  \abl{r}{t} -\left(\abl{r}{t}\right)^2,
\end{align}
we find
\begin{align}
  \abl{r}{t} &= c\left(1-\sqrt{\frac{r_s(t)}{r}}\right)\>.
               \label{eq:light_moving_out}
\end{align}
It can be shown analytically (setting $r=r_s+\delta$, $\delta\ll r_s$)
that for $r$ close enough to $r_s$ at some initial time, $r(t)$ will
in fact \bl{eventually} increase, allowing the light ray to escape, a
conclusion that is borne out by the numerics as we shall see
below. \bl{(Initially $r(t)$ decreases for $\delta<0$, but $r_s(t)$
  may decrease faster and as soon as $r(t)$ exceeds $r_s(t)$, it
  becomes an increasing function of time.)}
So $r_s(t)$ is not a true horizon anymore.

\section{Fate of infalling particles and outgoing light rays}
\label{sec:fate_infall}
Before discussing numerical results, let us estimate some time
scales. Taking $M_0=M_{\astrosun}=2\times 10^{30}\>$kg (i.e., one solar
mass) for an exemplary black hole, we find it to radiate thermally at
a temperature $T_{\text{BH}}=61.4\>$nK, according to
Eq.~\eqref{eq:hawking_temp}. This is less than a microkelvin, so to
radiate a solar mass away starting at that temperature will take a
huge amount of time. In reality, the temperature increases as the mass
decreases, therefore the process is self-accelerating. Nevertheless the life
time of a solar-mass black hole, evaluated from
Eq.~\eqref{eq:lifetime_BH}, is about
$6.7\times10^{74}\>$s$ = 2.1\times10^{67}\>$a, more than $10^{50}$
times the current age of the universe. Moreover, this estimate is valid only
for a black hole living in true vacuum. The temperature of the cosmic
microwave background (CMB) exceeds that of a black hole of mass
$M_{\astrosun}$, hence any existing black hole of that mass or higher
will grow by absorption of energy from that background rather than
shrink. Therefore, the process of evaporation will lead to an effective mass
decrease only after the CMB has cooled below $61.4\>$nK and from then
on take on the order of at least $10^{67}\>$a. Moreover, typical
stellar black holes have somewhat larger masses, so this is rather a
lower limit. A typical black hole with mass 10~$M_{\astrosun}$ will
need more than $10^{70}\>$a to decay, counting from the moment its
environment has a lower temperature than the black hole.

How much time does it take for a particle falling radially into a
black hole of mass $M_{\astrosun}$ from a distance of 1$\>\text{au}\; \bl{=1.496\times10^8}\>$\bl{km},
i.e., the distance of the earth from the sun? The (initial)
Schwarzschild radius of the sun is about 3$\>$km, and time dilation
does not kick in strongly before $r<10\, r_s$, even if we use $t_S$ to
measure time, so for most of the trip Newtonian physics provides a
good approximation. Kepler's third law gives us a result of
$1/(4\sqrt{2})\>\text{a}=0.177\>$a.\footnote{The trajectory of radial fall
  into the sun may be considered the first half of a degenerate Kepler
  ellipse with eccentricity 1, so the focus is the endpoint of the
  trajectory. The semimajor then is $r_0/2$, with $r_0$ the radius of
  Earth's orbit, making the ``period'' of that orbit equal to
  $\left(\frac{1}{2}\right)^{3/2}=1/2\sqrt{2}$ of the period of the
  Earth. So the time of fall to the center is $1/(4\sqrt{2})\>$a.}
Calculation of the four times $\tau_H$, $t_H$, $\tau_C$, and $t_C$
from Eqs.~\eqref{eq:tauH} through \eqref{eq:tC} gives 0.177$\>$a
for all four results, because their difference is not visible in the
first three significant digits. Time dilation remains negligible in
the PG time for this example and the first summand in each formula
dominates the others for $r_s\ll r_0$. Note that $\tau_C$ agrees
exactly with the Newtonian result, for obvious reasons
[Eq.~\eqref{eq:simil_newton} takes exactly the form of the
corresponding Newtonian equation of motion for constant $r_s$].

Since $0.177\>\text{a}\ll 10^{67}\>\text{a}$, the Schwarzschild radius
has not changed appreciably during the fall and the result calculated
for the static case should be an excellent approximation to what
happens on evaporation. This is of course very different from the
Aste-Trautmann model, where we have [$r_s=r_s(t_S)$]
\begin{align}
  c^2\left(\frac{\D\tau}{\D t_S}\right)^2 = \left(1-\frac{r_s}{r}\right)c^2-\left(1-\frac{r_s}{r}\right)^{-1}
  \left(\frac{\D r}{\D t_S}\right)^2
  \label{eq:fourvel_schwarzschild}
\end{align}
leading to a divergence of time dilation as $r_s$ is approached
($\D\tau/\D t_S \to 0$ and $\D r/\D t_S \to 0$, as the first term on
the right-hand side must exceed the second one outside $r_s$). The
corresponding equation for the model inspired by the PG variant of the
metric follows from \eqref{eq:fourvel}. It is
\begin{align}
  c^2\left(\frac{\D\tau}{\D t}\right)^2 = \left(1-\frac{r_s}{r}\right)c^2-2\sqrt{\frac{r_s}{r}} c \frac{\D r}{\D t}
  -\left(\frac{\D r}{\D t}\right)^2
  \label{eq:fourvel_PG}
\end{align}
and $\D \tau/\D t$ need not be small, because the second term on the
right-hand side is positive ($\D r/\D t<0$) and allows the left-hand
side to stay away from zero even as $r\to r_s$.

It is easy to derive
an upper bound of the time-dilation factor for $r>r_s$ from
Eq.~\eqref{eq:fourvel_schwarzschild}, using $\abs{\dot{r}/c}<1$, which yields
\begin{align}
  \abs{\frac{\D t_S}{\D\tau}} < \frac{\bl{\sqrt{2}}}{1-r_s/r}\>.
\end{align}
Therefore, the time dilation factor remains smaller than \bl{$2 \sqrt{2}$} for
$r>2r_s$. That is, down to 6 km from the center, there should not be
much of a difference between the behavior of the two models. Only then
the particle starts to seem to freeze at the horizon in the
Aste-Trautmann model for about $10^{67}$ years. Later, it would be
liberated, but may have been spaghettified
before,\footnote{\protect\url{https://en.wikipedia.org/wiki/Spaghettification}}
due to the smallness of the black hole preceding complete evaporation.

\begin{figure}[!htb]
  \vspace*{2mm}
  \centering
  \includegraphics[width=7.5cm]{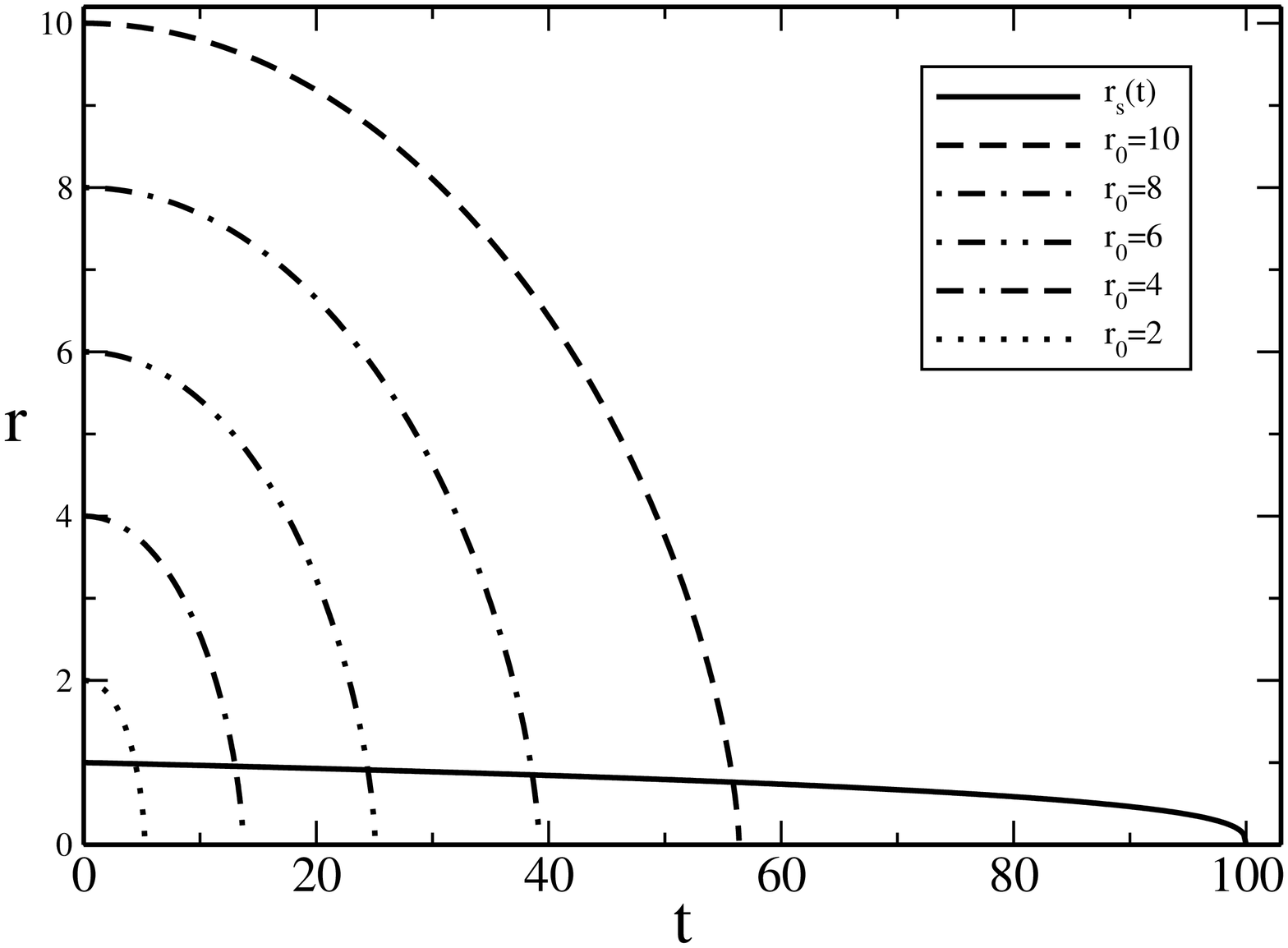}
  \caption{Trajectories of particles falling from different small multiples of $r_{s}(t=0)$ towards an
    evaporating black hole as modeled by the metric \eqref{eq:gp_metric_evap}. \label{fig:five_infalling_trajectories}}
\end{figure}

Numerical calculations were done in dimensionless units. The radial coordinate distance
to the spatial origin of the metric is measured in multiples of
$r_s$. As a time unit, we take a not too small fraction of $t_0$,
ranging between 0.01 and 1/5, in order to be able to see evaporation
on the time scale of our simulation. A few examples of the behavior of
trajectories of particles falling into an evaporating black hole
modeled as proposed here, are given in
Fig.~\ref{fig:five_infalling_trajectories}.

Note that by this choice of units, we are effectively considering
black holes with a very low initial mass $M_0$. Indeed, it is possible
to express the ratio between the characteristic times 
$\tau_C$,
 evaluated with the initial ($t=0$) Schwarzschild radius
$r_{s0}$,\footnote{The true proper time for falling to the singularity
  is of course somewhat longer in the model, as $r_s(t)$ decreases.} and $t_0$ as
\begin{align}
  \frac{\tau_C}{t_0} = \frac{1}{5120}\left(\frac{r_0}{r_{s0}}\right)^{3/2} \left(\frac{m_P}{M_0}\right)^2\>, 
\end{align}
where $m_P=\sqrt{\frac{\hbar c}{G}} = 2.176\times 10^{-8}\>$kg is the
Planck mass. To get this ratio close to one with
$r_0/r_{s0}\approx 300$ (which roughly cancels the denominator 5120),
we must have $M_0\approx m_P$. \bl{Let us assume $r_0= 1\>\text{au}$
  and check what mass $M_0$ is needed to obtain
  ${\tau_C}/{t_0}\approx 1$. It is useful to introduce
  $\tilde{m}_0 \equiv c^2 r_{0}/2G$ and to solve for $M_0/m_P$, which
  gives
\begin{align}
   \frac{M_0}{m_P} =  \left(\frac{t_0}{\tau_C}\right)^{2/7} \frac{1}{5120^{2/7}} \left(\frac{\tilde{m}_0}{m_P}\right)^{3/7}\>,
\end{align}
providing a mass of $M_0=7.06\times 10^{10}\>$kg (that corresponds to
the pretty small initial Schwarzschild radius
$r_{s0}=1,05\times10^{-16}\>$m) and a lifetime of
$t_0\approx \tau_C\approx 9.38\times 10^8\>$a.} Clearly, in scenarios
of this kind \bl{(with a free-fall time of $10^8\>$a for just one
  astronomical unit)} other masses \bl{that are present might} affect the falling
particle much stronger than the black hole (if we make the distance
larger to reach an initially more massive black hole near the end of
its life time, even distant stars will have to be taken into account
as trajectory-perturbing factors). Therefore, in most real-life
situations, we will have $\tau_C\ll t_0$. However, the only
qualitative difference between this case and $\tau_C\lesssim t_0$ is
that in the former case the curve describing $r_s(t)$ will be a
straight line parallel to the time axis, whereas in the latter we see
it approaching $r_s=0$.

What the calculation demonstrates is that letting the mass of the
evaporating black hole depend on a time variable that does not get
singular at $r_s$ and is meaningful also inside the Schwarzschild
radius leads to an infalling particle or observer being able to cross
$r_s$ without any problem, just as in the case of an eternal black
hole. There are good reasons to believe that this is the generic
behavior for any model using non-singular time coordinates to describe
evaporation. In particular, the enormous separation of time scales
between typical decay times of black holes and typical times to fall
in one, suggests that the case of an evaporating black hole is
indistinguishable from that of an eternal one, as far as crossing the
``horizon'' is concerned, in the large majority of cases.

\begin{figure}[!tb]
  \vspace*{1mm}
  \includegraphics[width=7.5cm]{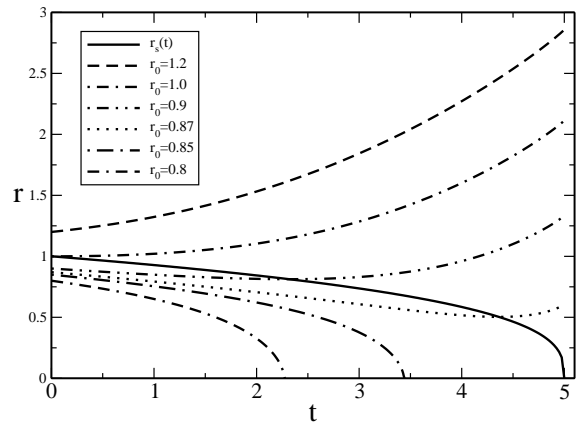}
  \caption{Light rays sent radially outward from different positions
    near the Schwarzschild radius or on
    it. \label{fig:some_outgoing_light_rays}}
\end{figure}

Let us now consider whether $r_s(t)$ is a horizon indeed. First, it is
clear that an outgoing radial light ray will momentarily hover at
$r=r_s(t)$ with zero coordinate speed, as
Eq.~\eqref{eq:gp_metric_evap} degenerates to $\D r(\D r + 2 c \D t)=0$
there, so the solutions for the local speed of light are $\D r/\D t=0$
(outgoing ray) and $\D r/\D t=-2c$ (ingoing ray). However, after
that instant $r_s$ has decreased a little, so the outgoing ray should
have a positive velocity. This is clearly true for $r=r_s$, hence such a
ray will escape. If we start our light ray \emph{inside} $r_s$,
however, escape may not be possible anymore. The numerical solution of
Eq.~\eqref{eq:light_moving_out} for several different initial
conditions is depicted together with $r_s(t)$ in
Fig.~\ref{fig:some_outgoing_light_rays}. The result demonstrates
unambiguously that there is a range of initial conditions extending
inside $r_s$ for which light can escape. On the other hand, light rays
starting too far inside $r_s(0)$ will still hit the singularity. This
means that a -- time-dependent -- event horizon continues to exist,
separating events from which there is no escape to future null
infinity and events from which a null ray can escape. However, the
radial coordinate of that horizon is smaller than $r_s(t)$.

\begin{figure}[!t]
  \vspace*{1mm}
  \centering
  \includegraphics[width=6.5cm]{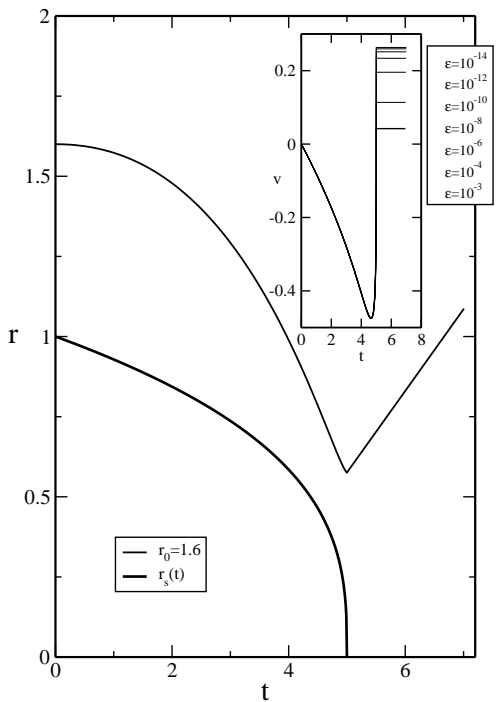} 
  \caption{Particle falling radially onto a black hole that manages
    to evaporate before being hit. The inset gives the velocity as a
    function of time for a series of diminishing parameters
    $\varepsilon$ to demonstrate convergence.\label{fig:evap_before_hit}}
\end{figure}

Finally, let us have a look at the case where an infalling particle
approaches a black hole in a way that it would hit the \bl{position of
  the} singularity only \emph{after} the evaporation time $t_0$
\bl{(i.e., when the singularity is gone)}. From the preceding discussion, it
should be clear given the vast disparity of time scales for falling
and evaporation that such a phenomenon, while not impossible, requires
quite some fine-tuning of the starting distance and time of the
particle and the initial mass of the black
hole. Figure~\ref{fig:evap_before_hit} displays an
example. Surprisingly, the particle starts getting repelled a short
time before $r_s$ turns zero and ends up with positive radial
velocity, i.e.~moving away from the center of the mass distribution of
the original black hole!

While this is very counterintuitive at first sight, it is definitely
predicted by the model. A look at Eq.~\eqref{eq:simil_newton} tells us
that the deviation from Newton's equation of motion, given by the
second term on the right-hand side, is positive, i.e.~repulsive, if
$r_s$ decreases with increasing time -- which is the case. 
Moreover, this term diverges as $r_s$ goes to zero. Since
$r_{s,t}\propto (t_0-t)^{-2/3}$ and
$1/\sqrt{r_s}\propto (t_0-t)^{-1/6}$, this divergence goes as
$(t_0-t)^{-5/6}$ and is integrable. ($\dot t$ remains bounded.)
Therefore, the velocity after $t=t_0$ will be finite. But it
may  be oriented away from the center, as the diverging repulsive
term eventually exceeds the leading attractive term that goes to zero as
$t\to t_0$. That the velocity is constant after complete evaporation
of the black hole, is expected -- the metric then is purely
Minkowskian. 

The unexpected behavior of the velocity of the infaller may be
interpreted in two different ways. In the first, its counterintuitive
aspects would be dismissed as a consequence of Newtonian thinking,
making us believe that disappearance of mass from the center should
only reduce gravitational attraction but not turn it into
repulsion. In the Einstein approach, the disappearance of mass
changes spacetime, i.e., it affects distances and time intervals. The
same radial coordinate corresponds to a smaller proper distance to the
center in the absence than in the presence of mass.\footnote{However,
  as long as there is a central singularity, such a statement can be
  made only for distances measured from a shell outside the center,
  because the proper distance to the latter is not defined.} If we
assume that the proper distance as a physical measure of remoteness is
continuous as central mass gets lost, this would suggest that the
radial coordinate must increase in order to avoid a change of proper
distance. This would result in an effective repulsion in terms of $r$.

The second way to interpret the repulsion would be to acknowledge that
this is just a toy model, in which mass does not disappear via a
well-defined physical process but rather by decree, so to speak. And
while the mass function of time may be realistic for times that are
small with respect to $t_0$, because there Hawking radiation can be
described perturbatively, it \bl{will probably} not describe things faithfully near
the point of complete evaporation, where a fully quantum mechanical
treatment is called for. This would then suggest that a more realistic
decay law would not allow for a divergent term in the equation of
motion. \bl{Rather,} it would  seem likely that the temporal decay happens much
more smoothly towards the end, a conclusion that has been reached in
Ref.~[\onlinecite{hiscock81a}] for different reasons.

\section{Conclusions}
\label{sec:conclus}

In this paper, it was shown that there is no \emph{a priori} reason to
believe that Hawking radiation will make a black hole evaporate from
under an observer, before she can fall in. While the argument is based
on a toy model, not treating the time-dependent metric to be expected
in the presence of Hawking radiation consistently, it has the
advantage of a time coordinate \bl{leading to a metric} that
\emph{(a)} does not become singular at the Schwarzschild radius
$r_s(t)$ and \emph{(b)} \bl{reduces to the}
Minkowski metric at infinity. The Aste-Trautmann model\cite{aste05} does
not have the first advantage, models based on Vaidya
metrics\cite{hiscock81a} do not exhibit the second. Therefore, the toy
model presented has the benefits of being both well-behaved and easily
interpretable. The time in which the mass function following from
Hawking radiation is usually expressed far from the black hole may be
made to coincide with the model global time.

As it turns out, in this model the Schwarzschild radius is as easily
crossed by a particle for an evaporating black hole as for an eternal
one, and the crossing happens in finite time. This behavior is
qualitatively different from that of the Aste-Trautmann model, which
uses as input a dependence of the mass function on the Schwarzschild
time. Conceptual problems of such an approach have been exposed.

\bl{What may be learned more generally from the results presented here?}

\bl{Evidently, the slowing-down of an observer falling towards the
  horizon of a static black hole may be largely considered an optical
  illusion. After all, local coordinate stationary observers will not
  perceive any slowing-down, whereas the perception of a distant
  observer is readily explained in terms of gravitational
  redshift. The divergence of the Schwarzschild time on approach to
  the horizon is explicable as the consequence of an unfortunate
  coordinate choice. It is not a fact of physics but rather one
  resulting from a convention. An appropriate coordinate
  transformation removes the divergence.}

\bl{However, things take a different twist, when time-dependent
  metrics are constructed by a generalization of either the
  Schwarzschild time or a non-singular time coordinate. The ensuing
  metrics are no longer related by a coordinate transformation but
  lead to physically different models. It then seems likely that the
  use of a coordinate that becomes singular at the Schwarzschild
  radius may produce spurious results.}

Since the two models are different, the fact that they oppose each
other even in a qualitative statement -- one predicts crossing of the
Schwarzschild radius while the other has the infaller freeze at the
apparent horizon until the black hole has evaporated -- is not
extremely surprising. Of course, at most one of the models can be
qualitatively right when compared with physical reality.

\bl{Finally, I would like to return to the question posed in the
  introduction whether a backreaction of the Hawking radiation might
  prevent formation of an event horizon or apparent horizon in the
  first place. Clearly, nothing rigorous can be said on the basis of
  the calculations presented here, as they do not consider
  backreaction. Nevertheless, our result may shed some light on the
  different approaches taken.}

\bl{At least some of the treatments answering the question in the
  affirmative\cite{gerlach76,mersini14} use a Schwarzschild-type time
  coordinate to describe the outer metric of a collapsing star. This
  coordinate becomes singular at an apparent horizon the moment it
  forms. It may then be difficult to separate a spurious coordinate
  effect from a true delay of horizon formation.\footnote{In
    Ref.~[\onlinecite{peiming16}], the absence of a horizon may be
    spurious due to the choice of a generalization of \emph{outgoing}
    Eddington-Finkelstein coordinates for the time dependent metric
    instead of the ingoing ones used in
    Ref.~[\onlinecite{hiscock81a}].}  On the other hand, the
  approaches answering the question in the negative
  \cite{parentani94,paranjape09,modak14} generally seem to use
  non-singular coordinates for the relevant calculations.}

   \bl{Obviously, results denying the formation
  of a black hole would be much more convincing, if they were entirely
  obtained using coordinates that remain regular across apparent
  horizons.\cite{shankaranarayanan02} }


\begin{thebibliography}{10}%
\makeatletter
\providecommand \@ifxundefined [1]{%
 \ifx #1\undefined \expandafter \@firstoftwo
 \else \expandafter \@secondoftwo
\fi
}%
\providecommand \@ifnum [1]{%
 \ifnum #1\expandafter \@firstoftwo
 \else \expandafter \@secondoftwo
\fi
}%
\providecommand \enquote [1]{``#1''}%
\providecommand \bibnamefont  [1]{#1}%
\providecommand \bibfnamefont [1]{#1}%
\providecommand \citenamefont [1]{#1}%
\providecommand\href[0]{\@sanitize\@href}%
\providecommand\@href[1]{\endgroup\@@startlink{#1}\endgroup\@@href}%
\providecommand\@@href[1]{#1\@@endlink}%
\providecommand \@sanitize [0]{\begingroup\catcode`\&12\catcode`\#12\relax}%
\@ifxundefined \pdfoutput {\@firstoftwo}{%
 \@ifnum{\z@=\pdfoutput}{\@firstoftwo}{\@secondoftwo}%
}{%
 \providecommand\@@startlink[1]{\leavevmode}%
 \providecommand\@@endlink[0]{}%
}{%
 \providecommand\@@startlink[1]{%
  \leavevmode
  \pdfstartlink
   attr{/Border[0 0 1 ]/H/I/C[0 1 1]}%
   user{/Subtype/Link/A<</Type/Action/S/URI/URI(#1)>>}%
  \relax
 }%
 \providecommand\@@endlink[0]{\pdfendlink}%
}%
\providecommand \url  [0]{\begingroup\@sanitize \@url }%
\providecommand \@url [1]{\endgroup\@href {#1}{\urlprefix}}%
\providecommand \urlprefix [0]{URL }%
\providecommand \Eprint[0]{\href }%
\@ifxundefined \urlstyle {%
  \providecommand \doi [1]{doi:\discretionary{}{}{}#1}%
}{%
  \providecommand \doi [0]{doi:\discretionary{}{}{}\begingroup
  \urlstyle{rm}\Url }%
}%
\providecommand \doibase [0]{http://dx.doi.org/}%
\providecommand \Doi[1]{\href{\doibase#1}}%
\providecommand \bibAnnote [3]{%
  \BibitemShut{#1}%
  \begin{quotation}\noindent
    \textsc{Key:}\ #2\\\textsc{Annotation:}\ #3%
  \end{quotation}%
}%
\providecommand \bibAnnoteFile [2]{%
  \IfFileExists{#2}{\bibAnnote {#1} {#2} {\input{#2}}}{}%
}%
\providecommand \typeout [0]{\immediate \write \m@ne }%
\providecommand \selectlanguage [0]{\@gobble}%
\providecommand \bibinfo [0]{\@secondoftwo}%
\providecommand \bibfield [0]{\@secondoftwo}%
\providecommand \translation [1]{[#1]}%
\providecommand \BibitemOpen[0]{}%
\providecommand \bibitemStop [0]{}%
\providecommand \bibitemNoStop [0]{.\EOS\space}%
\providecommand \EOS [0]{\spacefactor3000\relax}%
\providecommand \BibitemShut [1]{\csname bibitem#1\endcsname}%
\bibitem{aste05}%
  \BibitemOpen
  \bibfield{author}{%
  \bibinfo {author} {\bibfnamefont{A.}~\bibnamefont{Aste}}\ and\ \bibinfo
  {author} {\bibfnamefont{D.}~\bibnamefont{Trautmann}},\ }%
  \bibfield{title}{%
  \enquote{\bibinfo {title} {{Radial fall of a test particle onto an
  evaporating black hole}},}\ }%
  \bibfield{journal}{%
  \bibinfo {journal} {Canad. J. Phys.}\ }%
  \textbf{\bibinfo {volume} {83}},\ \bibinfo {pages} {1001--1006} (\bibinfo
  {year} {2005})%
  \bibAnnoteFile{NoStop}{aste05}%
\bibitem{vachaspati07}%
  \BibitemOpen
  \bibfield{author}{%
  \bibinfo {author} {\bibfnamefont{T.}~\bibnamefont{Vachaspati}}, \bibinfo
  {author} {\bibfnamefont{D.}~\bibnamefont{Stojkovic}},\ and\ \bibinfo {author}
  {\bibfnamefont{L.~M.}\ \bibnamefont{Krauss}},\ }%
  \bibfield{title}{%
  \enquote{\bibinfo {title} {{Observation of incipient black holes and the
  information loss problem}},}\ }%
  \bibfield{journal}{%
  \bibinfo {journal} {Phys. Rev. D}\ }%
  \textbf{\bibinfo {volume} {76}},\ \bibinfo {pages} {024005} (\bibinfo {year}
  {2007})%
  \bibAnnoteFile{NoStop}{vachaspati07}%
\bibitem{davis04}%
  \BibitemOpen
  \bibfield{author}{%
  \bibinfo {author} {\bibfnamefont{T.~M.}\ \bibnamefont{Davis}}\ and\ \bibinfo
  {author} {\bibfnamefont{C.~H.}\ \bibnamefont{Lineweaver}},\ }%
  \bibfield{title}{%
  \enquote{\bibinfo {title} {{Expanding Confusion: common misconceptions of
  cosmological horizons and the superluminal expansion of the universe}},}\ }%
  \bibfield{journal}{%
  \bibinfo {journal} {Publ. Astron. Soc. Australia}\ }%
  \textbf{\bibinfo {volume} {21}},\ \bibinfo {pages} {97--109} (\bibinfo {year}
  {2004})%
  \bibAnnoteFile{NoStop}{davis04}%
\bibitem{Note1}%
  \BibitemOpen
  \bibinfo {note} {A light signal sent from outside the Hubble sphere may reach
  us, due the the continuing expansion of the latter so it may eventually catch
  up with an inward-running light front. Once that light is inside the Hubble
  sphere, it will reach its center, in principle.}%
  \bibAnnoteFile{Stop}{Note1}%
\bl{\bibitem{gerlach76}%
  \BibitemOpen
  \bibfield{author}{%
  \bibinfo {author} {\bibfnamefont{U.H.}\ \bibnamefont{Gerlach}},\ }%
  \bibfield{title}{%
  \enquote{\bibinfo {title} {{The mechanism of blackbody radiation from an
  incipient black hole}},}\ }%
  \bibfield{journal}{%
  \bibinfo {journal} {Phys. Rev. D}\ }%
  \textbf{\bibinfo {volume} {14}},\ \bibinfo {pages} {1479--1508} (\bibinfo
  {year} {1976})%
  \bibAnnoteFile{NoStop}{gerlach76}%
\bibitem{mersini14}%
  \BibitemOpen
  \bibfield{author}{%
  \bibinfo {author} {\bibfnamefont{L.}~\bibnamefont{Mersini-Houghton}},\ }%
  \bibfield{title}{%
  \enquote{\bibinfo {title} {{Backreaction of Hawking radiation on a
  gravitationally collapsing star I: Black holes?}}.}\ }%
  \bibfield{journal}{%
  \bibinfo {journal} {Phys. Lett.}\ }%
  \textbf{\bibinfo {volume} {738}},\ \bibinfo {pages} {61--67} (\bibinfo {year}
  {2014})%
  \bibAnnoteFile{NoStop}{mersini14}%
\bibitem{peiming16}%
  \BibitemOpen
  \bibfield{author}{%
  \bibinfo {author} {\bibfnamefont{P.-M.}\ \bibnamefont{Ho}},\ }%
  \bibfield{title}{%
  \enquote{\bibinfo {title} {{The absence of horizon in black-hole
  formation}},}\ }%
  \bibfield{journal}{%
  \bibinfo {journal} {{Nuclear Physics B}}\ }%
  \textbf{\bibinfo {volume} {909}},\ \bibinfo {pages} {394--417} (\bibinfo
  {year} {2016})%
  \bibAnnoteFile{NoStop}{peiming16}%
\bibitem{parentani94}%
  \BibitemOpen
  \bibfield{author}{%
  \bibinfo {author} {\bibfnamefont{R.}~\bibnamefont{Parentani}}\ and\ \bibinfo
  {author} {\bibfnamefont{T.}~\bibnamefont{Piran}},\ }%
  \bibfield{title}{%
  \enquote{\bibinfo {title} {{The internal geometry of an evaporating black
  hole}},}\ }%
  \bibfield{journal}{%
  \bibinfo {journal} {Phys. Rev. Lett.}\ }%
  \textbf{\bibinfo {volume} {73}},\ \bibinfo {pages} {2805--2808} (\bibinfo
  {year} {1994})%
  \bibAnnoteFile{NoStop}{parentani94}%
\bibitem{paranjape09}%
  \BibitemOpen
  \bibfield{author}{%
  \bibinfo {author} {\bibfnamefont{A.}~\bibnamefont{Paranjape}}\ and\ \bibinfo
  {author} {\bibfnamefont{T.}~\bibnamefont{Padmanabhan}},\ }%
  \bibfield{title}{%
  \enquote{\bibinfo {title} {{Radiation from collapsing shells, semiclassical
  backreaction, and black hole formation}},}\ }%
  \bibfield{journal}{%
  \bibinfo {journal} {Phys. Rev. D}\ }%
  \textbf{\bibinfo {volume} {80}},\ \bibinfo {pages} {044011} (\bibinfo {year}
  {2009})%
  \bibAnnoteFile{NoStop}{paranjape09}%
\bibitem{modak14}%
  \BibitemOpen
  \bibfield{author}{%
  \bibinfo {author} {\bibfnamefont{S.K.}\ \bibnamefont{Modak}},\ }%
  \bibfield{title}{%
  \enquote{\bibinfo {title} {{Backreaction due to quantum tunneling and
  modification to the black hole evaporation process}},}\ }%
  \bibfield{journal}{%
  \bibinfo {journal} {Phys. Rev. D}\ }%
  \textbf{\bibinfo {volume} {90}},\ \bibinfo {pages} {044016} (\bibinfo {year}
  {2014})%
  \bibAnnoteFile{NoStop}{modak14}
  }%
\bibitem{einstein05}%
  \BibitemOpen
  \bibfield{author}{%
  \bibinfo {author} {\bibfnamefont{A.}~\bibnamefont{Einstein}},\ }%
  \bibfield{title}{%
  \enquote{\bibinfo {title} {{Zur Elektrodynamik bewegter K\"orper}},}\ }%
  \bibfield{journal}{%
  \bibinfo {journal} {Ann. Phys.}\ }%
  \textbf{\bibinfo {volume} {322}},\ \bibinfo {pages} {891--921} (\bibinfo
  {year} {1905}),\ \bibinfo {note} {{English translation in \emph{The Principle
  of Relativity} (Methuen, 1923, reprinted by Dover Publications, New York,
  1952), pp. 35 -- 65, ``On the electrodynamics of moving bodies''.}}%
  \bibAnnoteFile{Stop}{einstein05}%
\bibitem{einstein52}%
  \BibitemOpen
  \bibfield{author}{%
  \bibinfo {author} {\bibfnamefont{A.}~\bibnamefont{Einstein}},\ }%
  \emph{\bibinfo {title} {{Relativity: The Special and the General Theory}}},\
  \bibinfo {edition} {{15$^\text{th}$}}\ ed.\ (\bibinfo {publisher} {Crown},\
  \bibinfo {address} {New York},\ \bibinfo {year} {1952})%
  \bibAnnoteFile{NoStop}{einstein52}%
\bibitem{Note2}%
  \BibitemOpen
  \bibinfo {note} {This procedure makes the one-way speed of light equal to its
  round-trip velocity.}%
  \bibAnnoteFile{Stop}{Note2}%
\bibitem{rizzi04a}%
  \BibitemOpen
  \bibfield{author}{%
  \bibinfo {author} {\bibfnamefont{G.}~\bibnamefont{Rizzi}}, \bibinfo {author}
  {\bibfnamefont{M.~L.}\ \bibnamefont{Ruggiero}},\ and\ \bibinfo {author}
  {\bibfnamefont{A.}~\bibnamefont{Serafini}},\ }%
  \bibfield{title}{%
  \enquote{\bibinfo {title} {{Synchronization gauges and the principles of
  special relativity}},}\ }%
  \bibfield{journal}{%
  \bibinfo {journal} {Found. Phys.}\ }%
  \textbf{\bibinfo {volume} {34}},\ \bibinfo {pages} {1835--1887} (\bibinfo
  {year} {2004})%
  \bibAnnoteFile{NoStop}{rizzi04a}%
\bibitem{kassner12b}%
  \BibitemOpen
  \bibfield{author}{%
  \bibinfo {author} {\bibfnamefont{K.}~\bibnamefont{Kassner}},\ }%
  \bibfield{title}{%
  \enquote{\bibinfo {title} {{Ways to resolve Selleri's paradox}},}\ }%
  \bibfield{journal}{%
  \bibinfo {journal} {Am. J. Phys.}\ }%
  \textbf{\bibinfo {volume} {80}},\ \bibinfo {pages} {1061--1066} (\bibinfo
  {year} {2012})%
  \bibAnnoteFile{NoStop}{kassner12b}%
\bibitem{cranor99}%
  \BibitemOpen
  \bibfield{author}{%
  \bibinfo {author} {\bibfnamefont{M.~B.}\ \bibnamefont{Cranor}}, \bibinfo
  {author} {\bibfnamefont{E.~M.}\ \bibnamefont{Heider}},\ and\ \bibinfo
  {author} {\bibfnamefont{R.~H.}\ \bibnamefont{Price}},\ }%
  \bibfield{title}{%
  \enquote{\bibinfo {title} {{A circular twin paradox}},}\ }%
  \bibfield{journal}{%
  \bibinfo {journal} {Am. J. Phys.}\ }%
  \textbf{\bibinfo {volume} {68}},\ \bibinfo {pages} {1016--1020} (\bibinfo
  {year} {1999})%
  \bibAnnoteFile{NoStop}{cranor99}%
\bibitem{Note3}%
  \BibitemOpen
  \bibinfo {note} {To understand why, imagine that the arclength of a planar
  curve and a local normal coordinate are to be extended from a strip about the
  curve to the whole plane. If the curve is not straight, it will have
  osculating circles of finite radius. Clearly, is not possible to assign a
  unique set of coordinates to the center of such a circle. The normal
  coordinate at two different arclengths will point to that center, so it is
  not described by a single arclength coordinate. The situation gets worse for
  points even farther away from the curve considered.}%
  \bibAnnoteFile{Stop}{Note3}%
\bibitem{Note4}%
  \BibitemOpen
  \bibinfo {note} {Standard clocks are running at the rate of their proper
  time. The clocks of the Global Positioning System are nonstandard clocks
  similar to the ones showing Schwarzschild time, except that they run slow
  with respect to a local standard clock, in order to keep them synchronous
  with clocks on Earth, not at infinity.}%
  \bibAnnoteFile{Stop}{Note4}%
\bibitem{misner73}%
  \BibitemOpen
  \bibfield{author}{%
  \bibinfo {author} {\bibfnamefont{C.~W.}\ \bibnamefont{Misner}}, \bibinfo
  {author} {\bibfnamefont{K.~S.}\ \bibnamefont{Thorne}},\ and\ \bibinfo
  {author} {\bibfnamefont{J.~A.}\ \bibnamefont{Wheeler}},\ }%
  \emph{\bibinfo {title} {{Gravitation}}}\ (\bibinfo {publisher} {W. H.
  Freeman},\ \bibinfo {address} {New York},\ \bibinfo {year} {1973})%
  \bibAnnoteFile{NoStop}{misner73}%
\bibitem{Note5}%
  \BibitemOpen
  \bibinfo {note} {This rate is identical to that of a (standard) master clock
  at infinity.}%
  \bibAnnoteFile{Stop}{Note5}%
\bibitem{Note6}%
  \BibitemOpen
  \bibinfo {note} {Kruskal-Szekeres coordinates escape this conclusion by the
  fact that a coordinate stationary observer necessarily crosses the horizon.}%
  \bibAnnoteFile{Stop}{Note6}%
\bibitem{spivey15}%
  \BibitemOpen
  \bibfield{author}{%
  \bibinfo {author} {\bibfnamefont{R.}~\bibnamefont{Spivey}},\ }%
  \bibfield{title}{%
  \enquote{\bibinfo {title} {{Dispelling black hole pathologies through theory
  and observation}},}\ }%
  \bibfield{journal}{%
  \bibinfo {journal} {Progr. Phys.}\ }%
  \textbf{\bibinfo {volume} {11}},\ \bibinfo {pages} {321--329} (\bibinfo
  {year} {2015})%
  \bibAnnoteFile{NoStop}{spivey15}%
\bibitem{grib09}%
  \BibitemOpen
  \bibfield{author}{%
  \bibinfo {author} {\bibfnamefont{A.~A.}\ \bibnamefont{Grib}}\ and\ \bibinfo
  {author} {\bibfnamefont{Yu~V.}\ \bibnamefont{Pavlov}},\ }%
  \bibfield{title}{%
  \enquote{\bibinfo {title} {{Is it possible to see the infinite future of the
  Universe when falling into a black hole?}}.}\ }%
  \bibfield{journal}{%
  \bibinfo {journal} {Physics - Uspekhi}\ }%
  \textbf{\bibinfo {volume} {52}},\ \bibinfo {pages} {257--261} (\bibinfo
  {year} {2009})%
  \bibAnnoteFile{NoStop}{grib09}%
\bibitem{Note7}%
  \BibitemOpen
  \bibinfo {note} {See, for example, \\ \protect \url
  {https://en.wikipedia.org/wiki/Kruskal-Szekeres_coordinates}.}%
  \bibAnnoteFile{Stop}{Note7}%
\bibitem{Note8}%
  \BibitemOpen
  \bibinfo {note} {There is also an \protect \emph {antihorizon} in the
  diagram, having $t_S=-\infty $. It is the times between these limits $-\infty
  $ and $\infty $ that are missing from the horizons.}%
  \bibAnnoteFile{Stop}{Note8}%
\bibitem{painleve21}%
  \BibitemOpen
  \bibfield{author}{%
  \bibinfo {author} {\bibfnamefont{P.}~\bibnamefont{Painlevé}},\ }%
  \bibfield{title}{%
  \enquote{\bibinfo {title} {{La mécanique classique et la théorie de la
  relativité}},}\ }%
  \bibfield{journal}{%
  \bibinfo {journal} {{C. R. Acad. Sci. (Paris)}}\ }%
  \textbf{\bibinfo {volume} {173}},\ \bibinfo {pages} {677--680} (\bibinfo
  {year} {1921})%
  \bibAnnoteFile{NoStop}{painleve21}%
\bibitem{gullstrand22}%
  \BibitemOpen
  \bibfield{author}{%
  \bibinfo {author} {\bibfnamefont{A.}~\bibnamefont{Gullstrand}},\ }%
  \bibfield{title}{%
  \enquote{\bibinfo {title} {{Allgemeine Lösung des statischen
  Einkörperproblems in der Einsteinschen Gravitationstheorie}},}\ }%
  \bibfield{journal}{%
  \bibinfo {journal} {{Arkiv. Mat. Astron. Fys.}}\ }%
  \textbf{\bibinfo {volume} {16}},\ \bibinfo {pages} {1--15} (\bibinfo {year}
  {1922})%
  \bibAnnoteFile{NoStop}{gullstrand22}%
\bibitem{Note9}%
  \BibitemOpen
  \bibinfo {note} {We will assume this from now on, i.e., we set $c\protect
  \tmspace +\thinmuskip {.1667em} t_R= r_s \protect \qopname \relax o{ln}\left
  [(\protect \sqrt {R}+\protect \sqrt {r_s})/(\protect \sqrt {R}-\protect \sqrt
  {r_s})\right ]-2\protect \sqrt {R r_s}\protect \tmspace +\thinmuskip
  {.1667em}$.}%
  \bibAnnoteFile{Stop}{Note9}%
\bibitem{Note10}%
  \BibitemOpen
  \bibinfo {note} {This speed corresponds to the one an object falling from
  rest at infinity would have at $R$.}%
  \bibAnnoteFile{Stop}{Note10}%
\bibitem{Note11}%
  \BibitemOpen
  \bibinfo {note} {The light could be guided on the shell inside glass fibers,
  because for Einstein synchronization to work, we do not need the vacuum speed
  of light. All that is necessary is that the speed of the signal along the
  forward and backward directions of the fiber is the same.}%
  \bibAnnoteFile{Stop}{Note11}%
\bibitem{opatrny12}%
  \BibitemOpen
  \bibfield{author}{%
  \bibinfo {author} {\bibfnamefont{T.}~\bibnamefont{Opatrn\'y}}\ and\ \bibinfo
  {author} {\bibfnamefont{L}~\bibnamefont{Richterek}},\ }%
  \bibfield{title}{%
  \enquote{\bibinfo {title} {{Black hole heat engine}},}\ }%
  \bibfield{journal}{%
  \bibinfo {journal} {Am. J. Phys.}\ }%
  \textbf{\bibinfo {volume} {80}},\ \bibinfo {pages} {66--71} (\bibinfo {year}
  {2012})%
  \bibAnnoteFile{NoStop}{opatrny12}%
\bibitem{hiscock81a}%
  \BibitemOpen
  \bibfield{author}{%
  \bibinfo {author} {\bibfnamefont{W.~A.}\ \bibnamefont{Hiscock}},\ }%
  \bibfield{title}{%
  \enquote{\bibinfo {title} {{Models of evaporating black holes. I}},}\ }%
  \bibfield{journal}{%
  \bibinfo {journal} {Phys. Rev. D}\ }%
  \textbf{\bibinfo {volume} {23}},\ \bibinfo {pages} {2813--2822} (\bibinfo
  {year} {1981})%
  \bibAnnoteFile{NoStop}{hiscock81a}%
\bibitem{press86}%
  \BibitemOpen
  \bibfield{author}{%
  \bibinfo {author} {\bibfnamefont{W.~H.}\ \bibnamefont{Press}}, \bibinfo
  {author} {\bibfnamefont{B.~P.}\ \bibnamefont{Flannery}}, \bibinfo {author}
  {\bibfnamefont{S.~A.}\ \bibnamefont{Teukolsky}},\ and\ \bibinfo {author}
  {\bibfnamefont{W.~T.}\ \bibnamefont{Vetterling}},\ }%
  \emph{\bibinfo {title} {Numerical Recipes}}\ (\bibinfo {publisher} {Cambridge
  University Press},\ \bibinfo {address} {Cambridge},\ \bibinfo {year} {1986})%
  \bibAnnoteFile{NoStop}{press86}%
\bibitem{Note12}%
  \BibitemOpen
  \bibinfo {note} {The trajectory of radial fall into the sun may be considered
  the first half of a degenerate Kepler ellipse with eccentricity 1, so the
  focus is the endpoint of the trajectory. The semimajor then is $r_0/2$, with
  $r_0$ the radius of Earth's orbit, making the ``period'' of that orbit equal
  to $\left (\protect \frac {1}{2}\right )^{3/2}=1/2\protect \sqrt {2}$ of the
  period of the Earth. So the time of fall to the center is $1/(4\protect \sqrt
  {2})\mskip \medmuskip $a.}%
  \bibAnnoteFile{Stop}{Note12}%
\bibitem{Note13}%
  \BibitemOpen
  \bibinfo {note} {\protect \url
  {https://en.wikipedia.org/wiki/Spaghettification}}%
  \bibAnnoteFile{NoStop}{Note13}%
\bibitem{Note14}%
  \BibitemOpen
  \bibinfo {note} {The true proper time for falling to the singularity is of
  course somewhat longer in the model, as $r_s(t)$ decreases.}%
  \bibAnnoteFile{Stop}{Note14}%
\bibitem{Note15}%
  \BibitemOpen
  \bibinfo {note} {However, as long as there is a central singularity, such a
  statement can be made only for distances measured from a shell outside the
  center, because the proper distance to the latter is not defined.}%
  \bibAnnoteFile{Stop}{Note15}%
\bl{\bibitem{Note16}%
  \BibitemOpen
  \bibinfo {note} {In Ref.~[\protect \rev@citealpnum {peiming16}], the absence
  of a horizon may be spurious due to the choice of a generalization of
  \protect \emph {outgoing} Eddington-Finkelstein coordinates for the time
  dependent metric instead of the ingoing ones used in Ref.~[\protect
  \rev@citealpnum {hiscock81a}].}%
  \bibAnnoteFile{Stop}{Note16}%
\bibitem{shankaranarayanan02}%
  \BibitemOpen
  \bibfield{author}{%
  \bibinfo {author} {\bibfnamefont{S.}~\bibnamefont{Shankaranarayanan}},
  \bibinfo {author} {\bibfnamefont{T.}~\bibnamefont{Padmanabhan}},\ and\
  \bibinfo {author} {\bibfnamefont{K.}~\bibnamefont{Srinivasan}},\ }%
  \bibfield{title}{%
  \enquote{\bibinfo {title} {{Hawking radiation in different coordinate
  settings: complex paths approach}},}\ }%
  \bibfield{journal}{%
  \bibinfo {journal} {Class. Quant. Grav.}\ }%
  \textbf{\bibinfo {volume} {19}},\ \bibinfo {pages} {2671--2687} (\bibinfo
  {year} {2002})%
  \bibAnnoteFile{NoStop}{shankaranarayanan02}
}%
\end{thebibliography}
\end{document}